\documentclass[prd,showpacs,nofootinbib,preprint,10 pt]{revtex4}
\usepackage{amsmath,graphicx,color,xcolor,epsfig}

\def\Q{{\mathcal{Q}}}
\def\P{{\mathcal{P}}}

\begin{document}
\begin{titlepage}

\begin{flushright}
DESY 17-057\\
April 2017
\end{flushright}

\begin{center}
\setlength {\baselineskip}{0.3in} 
{\bf\Large\boldmath Mass spectrum of spin-1/2 pentaquarks with a $c\bar{c}$ component
and their anticipated discovery modes in $b$-baryon decays}\\[5mm]
\setlength {\baselineskip}{0.2in}
{\large  Ahmed Ali$^1$, Ishtiaq Ahmed$^{2}$, M. Jamil Aslam$^{3}$ and~ Abdur Rehman$^2$}\\[5mm]

$^1$~{\it Deutsches Elektronen-Synchrotron DESY, D-22607 Hamburg, Germany.}\\[5mm]

$^2$~{\it National Centre for Physics, Quaid-i-Azam University Campus,\\
             Islamabad 45320, Pakistan.}\\[5mm]
$^3$~{\it Physics Department, Quaid-i-Azam University,\\ Islamabad 45320, Pakistan.}\\[3cm] 
\end{center}

{\bf Abstract}\\[5mm] 
The LHCb discovery of the two baryonic states $P_c^+(4380)$ and $P_c^+(4450)$,
having $J^{P}=3/2^-$ and $J^{P}=5/2^+$, respectively, in the process $pp \to b \bar{b} \to \Lambda_b X$, followed by the decay $\Lambda_b \to J/\psi\; p\; K^-$,  has motivated a number of theoretical models. Interpreting them as compact $\{\bar{c}\; [c u] \; [ud];  L_{\mathcal{P}}=0,1\}$ objects, the mass spectroscopy of the $J^P=3/2^-$ and $J^P=5/2^+$ pentaquarks was worked out  by us for the pentaquarks in the $SU(3)_F$ multiplets, using an effective Hamiltonian based on constituent diquarks and quarks. Their possible discovery modes in $b$-baryon decays were also given using the heavy quark spin symmetry. In this paper, we calculate the mass spectrum of the hidden $c\bar{c}$ pentaquarks having
$J^{P}=\frac{1}{2}^\pm$ for the  $SU(3)_F$ multiplets and  their anticipated
discovery modes in $b$-baryon decays. Some of the $P_c^+(J^P=1/2^\pm$) pentaquarks, produced in the $\Lambda_{b}$ decays
may have their masses just below the $J/\psi\, p$ threshold, in which case they should be searched for
in the modes $ P_c^+(J^P=1/2^\pm) \to \eta_c\; p,\;\; \mu^+ \mu^- \;p,\;\;   e^+ e^-\; p$.

\end{titlepage}

\section{Introduction}
In 2015, LHCb  reported the first observation of two hidden charm pentaquark states $P_{c}^+(4380)$ and $P_c^+(4450)$ in the decay $\Lambda_b^0 \to J/\psi\; p\; K^-$ ~\cite{Aaij:2015tga}, having the masses
$4380 \pm 8 \pm 29$ MeV and $4449.8 \pm 1.7 \pm 2.5$ MeV, and widths $205 \pm 18 \pm 86$ MeV
and $39 \pm 5 \pm 19$ MeV, with the
preferred spin-parity assignments $J^{P}=\frac{3}{2}^-$ and $J^P=\frac{5}{2}^+$, respectively.
 These states have the quark composition $c\bar{c} uud$, and their masses 
  lie close to several (charm meson-baryon) thresholds. This has led to a number of theoretical proposals for their interpretation, which include  rescattering-induced kinematical effects \cite{rescattering}, open charm-baryon and  charm-meson bound states \cite{meson-boundstates}, and baryocharmonia \cite{Baryocharmonia}. They have also been interpreted as compact pentaquark hadrons with the internal structure organized as diquark-diquark-anti-charm quark~\cite{Maiani:2015vwa,compact-Pentaquark} or as diquark-triquark~\cite{Lebed:2015tna,Zhu:2015bba}.  

In an earlier paper~\cite{Ali:2016dkf} we followed the compact pentaquark interpretation, following the
basic idea that highly correlated diquarks play a key role in the physics of
multiquark states~\cite{Lipkin:1987sk,Jaffe:2003sg,Maiani:2004vq}. 
 The diquarks resulting from the
direct product $\tt 3 \otimes 3=\bar{3} \oplus 6$, are  either a color anti-triplet $\tt \bar{3}$ or a
color sextet $\tt 6$. Of these only the  color $\tt \bar{3}$ configuration is kept,
 as suggested by perturbative arguments. Both spin-1 and spin-0 diquarks are, however, allowed.
In the case of a diquark $[qq^\prime]$ consisting of two light quarks, the spin-0 diquarks are believed to be more tightly bound than the spin-1, but for
 the heavy-light diquarks, such as $[cq]$ or $[bq]$, this splitting is suppressed by $1/m_c$
for a $[cq]$ or by $1/m_b$ for a $[bq]$ diquark, and hence both spin-configurations are treated at par. 
For the pentaquarks, the mass spectrum  will depend upon how the five quarks, i.e., the 4 quarks and an antiquark, are dynamically  structured. A  diquark-triquark picture, in which the two observed pentaquarks consist of 
a rapidly separating pair
of a color-${\tt \bar{3}}$ $[cu]$ diquark and a color-${\tt 3}$ triquark $\bar{\theta}=\bar{c}[ud]$, has been
presented in~\cite{Lebed:2015tna}.

 In~\cite{Ali:2016dkf}, we  used the template in which the
5q baryons, such as the two $P_c$ states, are assumed to be four quarks, consisting of two highly
correlated diquark pairs, and an antiquark. For the  present discussion, it is an anti-charm quark $\bar{c}$ which
is correlated with the two
diquarks  $[cq]$ and $[q^{\prime}q^{\prime \prime}]$, where $q,q^{\prime},q^{\prime \prime}$ can be $u$ or $d$.
The tetraquark formed by the diquark-diquark $([cq]_{\tt \bar{3}}[q^{\prime}q^{\prime \prime}]_{\tt \bar{3}})$ is a color-triplet object,with orbital and spin quantum numbers,
 denoted by $L_{\Q\Q}$ and $S_{\Q\Q}$, 
which combines with the color-anti-triplet ${\tt \bar{3}}$ of the
 $\bar{c}$  to form an overall color-singlet pentaquark,
with the corresponding quantum numbers $L_{\P}$ and $S_{\P}$. (See,
Fig.~1 in~\cite{Ali:2016dkf}.)
An effective Hamiltonian based on this picture was constructed in~\cite{Ali:2016dkf}, extending the underlying tetraquark 
Hamiltonian developed for the $X,Y, Z$ states by Maiani {\it et al.}~\cite{Maiani:2004vq}. We
explained how the various input parameters  in this Hamiltonian were determined. Subsequently, we worked out the mass spectrum of the low-lying $S$- and $P$-wave 
pentaquark states, with a $c\bar{c}$ and three light quarks ($u$, $d$, $s$) in their Fock space,
but restricted ourselves to the $J^{P}=3/2^- $ and $J^{P}=5/2^+ $ states. The pentaquark states reported by the LHCb are produced in  $\Lambda_b^0$ decays, $\Lambda_b^0 \to \P^{+}\;K^{-}$, where $\P$ denotes a generic pentaquark state, a symbol we use subsequently in this work. 
In addition to the $\Lambda_b^0=(udb)$, which is the lightest of the $b$-baryons in which the light quark pair $ub$
has $j^P=0^+$, there are two others in this $SU(3)_F$ triplet with strangeness $S=-1$, 
$\Xi_b^0(5792)=(usb)$, having isospin $I=I_3=1/2$ and  $\Xi_b^-(5794)=(dsb)$, having isospin $I=-I_3=1/2$.
Likewise, there are six $b$-baryons with the light quark pair having $j^P=1^+$,  with $S=0$
 ($\Sigma_b^-=(ddb), \; \Sigma_b^0=(udb), \; \Sigma_b^+=(uub))$, $S=-1$ ($\Xi_b^\prime=(dsb),\;\Xi_b^{\prime 0}=(usb)$),  and one with $S=-2$ ($\Omega_b^-=(ssb) $.) These bottom baryon multiplets are shown in 
Fig. 2 of ref. \cite{Ali:2016dkf}.
We presented the discovery modes of the  $J^{P}=3/2^- $ and $J^{P}=5/2^+ $ pentaquarks
in $b$-baryon decays in~\cite{Ali:2016dkf}. In doing this, we assumed 
 heavy quark symmetry, i.e., for
 $m_b \gg \Lambda_{\rm QCD}$, $b$-quark becomes a static quark and the light diquark spin becomes a good quantum number, constraining the states which can otherwise be produced in these decays.    In particular, we found that in the diquark picture, one expects  a lower-mass
  $J^P=\frac{3}{2}^-$ pentaquark state  with the quantum numbers $\{\bar{c} [cu]_{s=1} [ud]_{s=0}; L_{\mathcal{P}}=0, J^{\rm P}=\frac{3}{2}^- \}$, which has the correct light diquark spin  to be produced in the decay
$\Lambda_b^0 \to J/\psi\; p\; K^-$,  compatible with the heavy quark symmetry. We estimated
its mass in the range 4110 - 4130 MeV, and suggested to search for the lower mass $P_c^+ (J^P = \frac{3}{2}^-)$ state decaying into $J/\psi\; p$  in the  LHCb data
on $\Lambda_b^0 \to J/\psi\; p K^-$. 

In this paper, we extend
the mass spectrum calculation to the $J^{P}=1/2^\pm $ pentaquark case. The 
$S$ ($P$)- wave pentaquark states are called $\mathcal{P}_{X_{i}}$ ($\mathcal{P}_{Y_{i}}$),
and their spin- and orbital angular momentum quantum numbers are given in Tables \ref{Table I} and \ref{Table I-a}. There
are five $S$-wave pentaquark states with $J^P=\frac{1}{2}^-$ , four $S$-wave pentaquark states
with  $J^P=\frac{3}{2}^-$, and nine $P$-wave states with
$J^P=\frac{1}{2}^+$. The input constituent quark masses and inter-diquark and intra-diquark spin-spin couplings are given in
Tables II and III of ref. \cite{Ali:2016dkf} and the quark flavors for the pentaquark multiplets are given in Table \ref{Table III}. The mass term $\Delta M$ that arises from different spin-spin interactions is given in Table \ref{TableM15}. The masses of
the five $S$-wave pentaquarks with $J^P=\frac{1}{2}^-$ are given in Table \ref{TableIV-1}, and the
masses of the nine $P$-wave pentaquarks with $J^P=\frac{1}{2}^+$ are given in Tables \ref{TableIV-2} and \ref{TableIV-3}.
We also work out the discovery modes of the pentaquarks with $J^P=\frac{1}{2}^\pm$
in  various $b$-baryon decays. In doing this, we have used $SU(3)_F$ and heavy quark
symmetries, discussed in~\cite{Ali:2016dkf}. We find that some of the lowest-lying pentaquarks
may have their masses below the threshold to decay into a $J/\psi p$ (and similar thresholds in
other pentaquarks). In this case, the discovery modes are expected to be
 $\eta_c\; p,\;\; \mu^+ \mu^- \;p$ and $  e^+ e^-\; p$.
 Estimate of the ratios of decay widths $\Gamma(\mathcal{B}(\mathcal{C}) \to \mathcal{P}^{1/2^{-}}\mathcal{M})/\Gamma(\Lambda^{0}_b \to P_{p}^{{\{Y_2\}}_{c_1}}K^{-})$ for $\Delta S = 1$ transitions are given in Table \ref{Relative-Rates11}. Here 
 $\mathcal{M}$ is the lightest pseudoscalar meson nonet and 
 $P_{p}^{{\{Y_2\}}_{c_1}}$ is the state with the mass $4450$ MeV and $J^{P} = \frac{5}{2}^{+}$, measured recently 
 by the LHCb  \cite{Aaij:2015tga}. The symbols $\mathcal{B} $ and $\mathcal{C}$ represent the flavor-antitriplet $b$-baryons with the light-quark spin $j^P=0^+$, and the flavor-sextet $b$-baryons with $j^P=1^+$,
respectively. The corresponding $\Delta S = 0$ transitions are given in Table \ref{Relative-Rates12}. Those involving
pentaquarks with $\mathcal{P}^{1/2^{+}}$ are given in Tables \ref{Relative-Rates13} (for $\Delta S = 1$ transitions),
and in Table \ref{Relative-Rates14} (for $\Delta S = 0$ transitions).

 This paper is organized as follows. In section \ref{spectrum-sec}, we work out the pentaquark mass spectrum with a
 hidden $c\bar{c}$ component  and having $J^{P}=1/2^\pm $, using the effective 
 Hamiltonian~\cite{Ali:2016dkf}. Numerical estimates of the pentaquark masses are given in section III. 
 In section IV, we present the weak decays of the $b$-baryons,
into a pseudoscalar meson and a pentaquark state with $\mathcal{P}^{1/2^{\pm}}$.
  We conclude in section V with a discussion of the $\boldmath{J = \frac{1}{2}}$ pentaquark decays
   and the various corresponding meson-baryon thresholds.
%
\section{Effective Hamiltonian framework for Pentaquark spectrum}\label{spectrum-sec}
Assuming that the underlying structure of the pentaquarks is given by $\bar{c}[cq][q^{\prime}q^{\prime \prime}]$, we calculate the mass spectrum of these states by extending the
effective Hamiltonian proposed for the tetraquark spectroscopy by Maiani
\textit{et al.}~\cite{Maiani:2014aja}. The resulting Hamiltonian for pentaquarks is \cite{Ali:2016dkf}
\begin{equation}
H=H_{[\mathcal{Q}\mathcal{Q}^{\prime}]} + H_{\bar{c}[\mathcal{Q}\mathcal{Q}^\prime]} + H_{S_{\P} L_{\P}} + H_{L_{\P} L_{\P}}\label{main-Hamiltion} ,
\end{equation}
where $\mathcal{Q}$ and $\mathcal{Q}
^{\prime}$ denote the diquarks $[cq]$ and $[q^{\prime}q^{\prime \prime}]$ having masses $m_{\mathcal{Q}}$ and $m_{\mathcal{Q}^{\prime}}$,
respectively.  The individual terms in the Hamiltonian \eqref{main-Hamiltion} are
\begin{eqnarray}
\begin{array}{rclrclrcl}
H_{[\mathcal{Q}\mathcal{Q}^{\prime}]} &=&m_{\mathcal{Q}}+m_{\mathcal{Q}^{\prime}}+H_{SS}(\mathcal{Q}\mathcal{Q}^\prime) + H_{SL}(\mathcal{Q}\mathcal{Q}^\prime)+H_{LL}(\mathcal{Q}\mathcal{Q}^\prime),  &
H_{S_{\mathcal{P}}L_{\mathcal{P}}} &=&2A_{\mathcal{P}}(\mathbf{S}_{\mathcal{P}}\cdot
\mathbf{L_{\mathcal{P}}}), &
H_{L_{\mathcal{P}}L_{\mathcal{P}}}&=&B_{\mathcal{P}}\frac{L_{\mathcal{P}}(L_{\mathcal{P}}+1)}{2},
\end{array} 
\end{eqnarray}
with
\begin{eqnarray}
H_{SS}(\mathcal{Q}\mathcal{Q}^\prime) &=& 2(\mathcal{K}_{cq})_{\bar{3}}(\mathbf{S}_{c}\cdot
\mathbf{S}_{q})+2(\mathcal{K}_{q^{\prime}q^{\prime \prime}})_{\bar{3}}(\mathbf{S}_{q^{\prime}}\cdot
\mathbf{S}_{q^{\prime \prime}}) \label{02}, \\
H_{\bar{c}[\mathcal{Q}\mathcal{Q}^{\prime}]}&=& m_c+2\,\mathcal{K}_{\bar{c}c}(\mathbf{S}_{\bar{c}}\cdot
\mathbf{S}_{c})+2\mathcal{K}_{\bar{c}q}(\mathbf{S}_{\bar{c}}\cdot
\mathbf{S}_{q}) +2\mathcal{K}_{\bar{c}q^{\prime}}(\mathbf{S}_{\bar{c}}\cdot
\mathbf{S}_{q^{\prime}})+2\mathcal{K}_{\bar{c}q^{\prime \prime}}(\mathbf{S}_{\bar{c}}\cdot
\mathbf{S}_{q^{\prime \prime}}). \label{02-1}
\end{eqnarray}
$L_{\P}$ and $S_{\P}$ are the orbital
angular momentum and the spin of the pentaquark state, and the quantities $A_{\mathcal{P}}$ and  $B_{\mathcal{P}}$ indicate the strength of their
spin-orbit  and orbital angular momentum couplings, respectively. The values of diquark masses and that of $A_{\mathcal{P}}$ and  $B_{\mathcal{P}}$ are given in ref. \cite{Ali:2016dkf}. The parameters
  $(\mathcal{K}_{cq})_{\bar{3}%
}$ and $(\mathcal{K}_{q^{\prime}q^{\prime \prime}})_{\bar{3}}$ correspond to the couplings of spin-spin interactions between 
the quarks within the diquarks. The other terms that correspond to the spin and orbital angular momentum couplings of the tetraquark are
\begin{equation}
H_{SL}(\Q\Q^\prime)  = 2 A_{\Q\Q^\prime} \mathbf{S}_{\Q \Q^\prime}\cdot \mathbf{L}_{\Q\Q^\prime}
,\;\;  H_{LL} = B_{\Q \Q^\prime} \frac{L_{\Q\Q^\prime}(L_{\Q\Q^\prime}+1)}{2}.
\label{03}
\end{equation}
In Model II proposed by Maiani \textit{et al.}~\cite{Maiani:2014aja}, it is assumed that the quarks 
in a diquark are tightly bound, and only their spin-spin coupling is kept, whereas 
in their earlier model~\cite{Maiani:2004vq} (called Model I), the couplings among the quarks of the two diquarks were also included.  This amounts to 
 adding four additional spin-spin terms in the $H_{SS}(\Q\Q^\prime)$ part of Hamiltonian given in Eq. (\ref{02}).
\begin{equation}
H_{SS}(\Q\Q^\prime) = 2(\mathcal{K}_{cq})_{\bar{3}}(\mathbf{S}_{c}\cdot
\mathbf{S}_{q})+2(\mathcal{K}_{q^{\prime}q^{\prime \prime}})_{\bar{3}}(\mathbf{S}_{q^{\prime}}\cdot
\mathbf{S}_{q^{\prime \prime}})+2(\mathcal{K}_{cq^\prime})_{\bar{3}}(\mathbf{S}_{c}\cdot
\mathbf{S}_{q^\prime})+2(\mathcal{K}_{cq^{\prime \prime}})_{\bar{3}}(\mathbf{S}_{c}\cdot
\mathbf{S}_{q^{\prime \prime}})+2(\mathcal{K}_{qq^\prime})_{\bar{3}}(\mathbf{S}_{q}\cdot
\mathbf{S}_{q^\prime})+2(\mathcal{K}_{qq^{\prime \prime}})_{\bar{3}}(\mathbf{S}_{q}\cdot
\mathbf{S}_{q^{\prime \prime}}). \label{02-2}
\end{equation}
We have taken all the couplings to be positive.

The mass formula for the pentaquark state which contains a ground state tetraquark ($%
L_{{\mathcal{Q}}{\mathcal{Q}}^{\prime }}=0)$ can be determined by the following formula
\begin{equation}
M=M_{0}+\frac{B_{\mathcal{P}}}{2}L_{\mathcal{P}}(L_{\mathcal{P}}+1)+2A_{%
\mathcal{P}}\frac{J_{\mathcal{P}}(J_{\mathcal{P}}+1)-L_{\mathcal{P}}(L_{%
\mathcal{P}}+1)-S_{\mathcal{P}}(S_{\mathcal{P}}+1)}{2}+\Delta M,
\label{mass-formula}
\end{equation}%
where $M_{0}=m_{\mathcal{Q}}+m_{\mathcal{Q}^{\prime }}+m_{c}$ and $\Delta M$ is the mass term
that emerges from different spin-spin interactions. 

We have classified the states in terms
of the diquarks spins, $S_{\mathcal{Q}}$ and $S_{\mathcal{Q}^{\prime }}$, the spin of the anti-charm
quark $S_{\bar{c}}=1/2$, the orbital angular momentum $L_{\mathcal{P}}$, and the total $J$ of the pentaquark $|S_{\mathcal{Q}},S_{\mathcal{Q}^{\prime }},S_{\bar{c}}, L_{\mathcal{P}}; J \rangle$:
\begin{eqnarray}
|0_{\mathcal{Q}},0_{\mathcal{Q}^{\prime }},\frac{1}{2}_{\bar{c}}, L_{\mathcal{P}};\frac{1}{2}\rangle _{1} &=&%
\frac{1}{2}[\left( \uparrow \right) _{c}\left( \downarrow \right)
_{q}-\left( \downarrow \right) _{c}\left( \uparrow \right) _{q}][\left(
\uparrow \right) _{q^{\prime }}\left( \downarrow \right) _{q^{\prime \prime
}} - \left(\downarrow \right) _{q^{\prime }}\left( \uparrow \right) _{q^{\prime \prime
}}]\left( \uparrow \right) _{\bar{c}}  \notag \\
|0_{\mathcal{Q}},1_{\mathcal{Q}^{\prime }},\frac{1}{2}_{\bar{c}}, L_{\mathcal{P}};\frac{1}{2}\rangle _{2} &=&%
\frac{1}{\sqrt{3}}[\left( \uparrow \right) _{c}\left( \downarrow \right)
_{q}-\left( \downarrow \right) _{c}\left( \uparrow \right) _{q}]\{\left(
\uparrow \right) _{q^{\prime }}\left( \uparrow \right) _{q^{\prime \prime}}\left( \downarrow \right) _{\bar{c}}-\frac{1}{2}[\left(
\uparrow \right) _{q^{\prime }}\left( \downarrow \right) _{q^{\prime \prime
}} + \left(\downarrow \right) _{q^{\prime }}\left( \uparrow \right) _{q^{\prime \prime
}}]\left( \uparrow \right) _{\bar{c}}\}  \notag \\
|1_{\mathcal{Q}},0_{\mathcal{Q}^{\prime }},\frac{1}{2}_{\bar{c}}, L_{\mathcal{P}};\frac{1}{2}\rangle _{3} &=&%
\frac{1}{\sqrt{3}}[\left(
\uparrow \right) _{q^{\prime }}\left( \downarrow \right) _{q^{\prime \prime
}} - \left(\downarrow \right) _{q^{\prime }}\left( \uparrow \right) _{q^{\prime \prime
}}]\{\left(\uparrow \right) _{c}\left( \uparrow \right) _{q}\left( \downarrow \right) _{\bar{c}}-\frac{1}{2}[\left( \uparrow \right) _{c}\left( \downarrow \right)_{q}+\left( \downarrow \right) _{c}\left( \uparrow \right) _{q}]\left( \uparrow \right) _{\bar{c}}\}  \notag \\
|1_{\mathcal{Q}},1_{\mathcal{Q}^{\prime }},\frac{1}{2}_{\bar{c}}, L_{\mathcal{P}};\frac{1}{2}\rangle _{4} &=&%
\frac{1}{3}\left( \uparrow \right) _{c}\left( \uparrow \right)_{q}\{[\left(
\uparrow \right) _{q^{\prime }}\left( \downarrow \right) _{q^{\prime \prime
}} + \left(\downarrow \right) _{q^{\prime }}\left( \uparrow \right) _{q^{\prime \prime
}}]\left( \downarrow \right) _{\bar{c}}-2\left(\downarrow \right) _{q^{\prime }}\left( \downarrow \right) _{q^{\prime \prime}}\left( \uparrow \right) _{\bar{c}}\}\notag \\ 
&&-\frac{1}{6}[\left( \uparrow \right) _{c}\left( \downarrow \right)
_{q}+\left( \downarrow \right) _{c}\left( \uparrow \right) _{q}]\{2\left(\uparrow \right) _{q^{\prime }}\left( \uparrow \right) _{q^{\prime \prime}}\left( \downarrow \right) _{\bar{c}}-[\left(\uparrow \right) _{q^{\prime }}\left( \downarrow \right) _{q^{\prime \prime}} + \left(\downarrow \right) _{q^{\prime }}\left( \uparrow \right) _{q^{\prime \prime}}]\left( \uparrow \right) _{\bar{c}}\} \notag \\
|1_{\mathcal{Q}},1_{\mathcal{Q}^{\prime }},\frac{1}{2}_{\bar{c}}, L_{\mathcal{P}};\frac{1}{2}\rangle _{5} &=&%
\frac{1}{\sqrt{2}}\left( \downarrow \right)_{c}\left( \downarrow \right)_{q} \left(\uparrow \right) _{q^{\prime }}\left( \uparrow \right) _{q^{\prime \prime}}\left( \uparrow \right) _{\bar{c}}+\frac{1}{3\sqrt{2}}\left( \uparrow \right)_{c}\left( \uparrow \right)_{q}\{[\left(\uparrow \right) _{q^{\prime }}\left( \downarrow \right) _{q^{\prime \prime
}} + \left(\downarrow \right) _{q^{\prime }}\left( \uparrow \right) _{q^{\prime \prime
}}]\left( \downarrow \right) _{\bar{c}}+\left(\downarrow \right) _{q^{\prime }}\left( \downarrow \right) _{q^{\prime \prime}}\left( \uparrow \right) _{\bar{c}}\}\notag \\
&&-\frac{1}{3\sqrt{2}}[\left( \uparrow \right) _{c}\left( \downarrow \right)
_{q}+\left( \downarrow \right) _{c}\left( \uparrow \right) _{q}]\{\left(\uparrow \right) _{q^{\prime }}\left( \uparrow \right) _{q^{\prime \prime}}\left( \downarrow \right) _{\bar{c}}+[\left(
\uparrow \right) _{q^{\prime }}\left( \downarrow \right) _{q^{\prime \prime
}} + \left(\downarrow \right) _{q^{\prime }}\left( \uparrow \right) _{q^{\prime \prime
}}]\left( \uparrow \right) _{\bar{c}}\}. \label{Spinj1by2}
\end{eqnarray}%
Using $L_{\mathcal{P}} = 0$ and $L_{\mathcal{P}} = 1$ in the basis defined in Eq. (\ref{Spinj1by2}), we have 
five $S$-wave pentaquark states for $J^{P}=\frac{1}{2}^{-}$ and five $P$-wave states with $J^{P}=\frac{1}{2}^{+}$, respectively. In all these states the net spin of pentaquark state $S_{\mathcal{P}} = \frac{1}{2}$.

Using the states given in Eq.~(\ref{Spinj1by2}), the
mass splitting matrix $\Delta M$ is a symmetric
$ (5 \times 5)$ matrix. Denoting its elements by $m_{ij}$ $(i , j=1,\cdots,5)$,
their diagonal entries can be written in terms of the spin-spin couplings as follows:
\begin{eqnarray}
m _{11} &=&-\frac{3}{4}((\mathcal{K}%
_{q^{\prime }q^{\prime \prime }})_{\bar{3}}+(\mathcal{K}_{cq})_{\bar{3}}), \; 
m _{22} =\frac{1}{4}(-3(\mathcal{K}_{cq})_{\bar{3}}+(\mathcal{K}_{q^{\prime }q^{\prime \prime }})_{%
\bar{3}}-5(\mathcal{K}_{\bar{c}q^{\prime}}+\mathcal{K}_{\bar{c}q^{\prime \prime }})),  \notag   \\
m _{33} &=&\frac{1}{4}(-3(%
\mathcal{K}_{q^{\prime}q^{\prime \prime }})_{\bar{3}}+(\mathcal{K}_{cq})_{%
\bar{3}}-\frac{5}{3}(\mathcal{K}_{\bar{c}q^{\prime}}+\mathcal{K}_{\bar{c}c})),   \label{matrix-entires}\\
m _{44} &=&\frac{1}{4}((\mathcal{K}_{cq})_{\bar{3}}+(\mathcal{K}%
_{q^{\prime }q^{\prime \prime }})_{\bar{3}})-\frac{1}{36}(\mathcal{K}_{\bar{c}q}+\mathcal{K}_{\bar{c}c})-
\frac{2}{3}((\mathcal{K}_{qq^{\prime }})_{%
\bar{3}}+(\mathcal{K}_{qq^{\prime \prime }})_{\bar{3}})-\frac{1}{2}(\mathcal{K}_{\bar{c}%
q^{\prime }}+\mathcal{K}_{\bar{c}q^{\prime \prime }}), \notag \\
m _{55} &=&-\frac{4}{9}(\mathcal{K}_{\bar{c}q}+\mathcal{K}_{\bar{c}c})-\frac{5}{12}((\mathcal{K}_{cq^{\prime}})_{\bar{3}}+(\mathcal{K}%
_{c q^{\prime \prime }})_{\bar{3}}+(\mathcal{K}_{qq^{\prime}})_{\bar{3}}+(\mathcal{K}_{qq^{\prime \prime }})_{\bar{3}})+\frac{1}{6}(\mathcal{K}_{\bar{c}q^{\prime}}+\mathcal{K}_{\bar{c}q^{\prime\prime}})+\frac{47}{72}((\mathcal{K}_{cq})_{\bar{3}}+(\mathcal{K}%
_{q^{\prime }q^{\prime \prime }})_{\bar{3}}).\notag
\end{eqnarray}
\vspace{-0.9cm}
Similarly, the off-diagonal entries take the form

\begin{eqnarray}
m _{12} &=& -\frac{3}{4\sqrt{3}}(\mathcal{K}_{\bar{c}q^{\prime }}-\mathcal{K}_{\bar{c}q^{\prime \prime }}), \quad \; 
m _{13} =-\frac{3}{4\sqrt{3}}(\mathcal{K}_{\bar{c}c}-\mathcal{K}_{\bar{c}q}), \quad \;
m _{14} =\frac{1}{4}((\mathcal{K}_{cq^{\prime
}})_{\bar{3}}-(\mathcal{K}_{cq^{\prime \prime }})_{\bar{3}}+(\mathcal{K}%
_{qq^{\prime }})_{\bar{3}}-(\mathcal{K}_{qq^{\prime \prime }})_{\bar{3}}),  \notag \\
m _{15} &=&-\frac{1}{2\sqrt{2}}((\mathcal{K}_{cq^{\prime
}})_{\bar{3}}-(\mathcal{K}_{cq^{\prime \prime }})_{\bar{3}}+(\mathcal{K}%
_{qq^{\prime }})_{\bar{3}}-(\mathcal{K}_{qq^{\prime \prime }})_{\bar{3}}),\;   \quad\quad\;
m _{23} =\frac{1}{4}((\mathcal{K}%
_{cq^{\prime }})_{\bar{3}}+(\mathcal{K}_{cq^{\prime \prime }})_{%
\bar{3}}-(\mathcal{K}_{qq^{\prime }})_{\bar{3}}-(%
\mathcal{K}_{qq^{\prime \prime }})_{\bar{3}}),  \notag \\
m _{24} &=&-\frac{5}{6\sqrt{3}}((\mathcal{K}_{cq^{\prime
}})_{\bar{3}}+(\mathcal{K}_{cq^{\prime \prime }})_{\bar{3}}-(\mathcal{K}%
_{qq^{\prime }})_{\bar{3}}-(\mathcal{K}_{qq^{\prime \prime }})_{\bar{3}}), 
\notag \\
m _{25} &=&-\frac{1}{2\sqrt{6}}(2\mathcal{K}_{\bar{c}c}-2\mathcal{K}_{\bar{c}q}-(\mathcal{K}_{cq^{\prime
}})_{\bar{3}}-(\mathcal{K}_{cq^{\prime \prime }})_{\bar{3}}+(\mathcal{K}%
_{qq^{\prime }})_{\bar{3}}+(\mathcal{K}_{qq^{\prime \prime }})_{\bar{3}}),\notag \\
m _{34} &=&-\frac{5}{6\sqrt{3}}((\mathcal{K}_{cq^{\prime }})_{\bar{3}}+%
(\mathcal{K}_{cq^{\prime \prime }})_{\bar{3}}-(%
\mathcal{K}_{qq^{\prime }})_{\bar{3}}-(\mathcal{K}_{qq^{\prime
\prime }}))_{\bar{3}}),  \notag \\
m _{35} &=&\frac{1}{2\sqrt{6}}(2\mathcal{K}_{\bar{c}q^{\prime}}-2\mathcal{K}_{\bar{c}q^{\prime \prime}}+(\mathcal{K}_{cq^{\prime
}})_{\bar{3}}+(\mathcal{K}_{cq^{\prime \prime }})_{\bar{3}}-(\mathcal{K}%
_{qq^{\prime }})_{\bar{3}}+(\mathcal{K}_{qq^{\prime \prime }})_{\bar{3}}), 
\notag \\
m _{45} &=&\frac{1}{\sqrt{2}}(-\frac{2}{9}(\mathcal{K}_{\bar{c}q}+\mathcal{K}_{\bar{c}c})+\frac{1}{6}((\mathcal{K}_{cq^{\prime}})_{\bar{3}}+(\mathcal{K}%
_{c q^{\prime \prime }})_{\bar{3}})+\frac{1}{6}((\mathcal{K}_{qq^{\prime}})_{\bar{3}}+(\mathcal{K}_{qq^{\prime \prime }})_{\bar{3}})).\label{off-diag-matrix-entries}
\end{eqnarray}
From the  above expressions, given for Model I, one obtains 
the expressions for  Model II~\cite{Maiani:2014aja}
 by setting the couplings $(\mathcal{K}_{cq^{\prime }})_{\bar{3}}$, $(\mathcal{K}%
_{cq^{\prime \prime }})_{\bar{3}}$, $(\mathcal{K}_{qq^{\prime }})_{\bar{3}}$ to zero.
In Table~\ref{Table I}, $\Delta M_{i}$  ($i$ runs from $1$ to $5$) are the mass splitting terms that arise after the
diagonalization of the $5 \times 5$ matrix whose entries are given in Eqs.~(\ref{matrix-entires}) and (\ref{off-diag-matrix-entries}).
In addition to this, there are four $J^{p} =\frac{1}{2}^{+}$ states, which result on combining the spin $S_{\mathcal{P}}=\frac{3}{2}$ with the orbital angular momentum $L_{\mathcal{P}} = 1$. In these state the value of $A_{\mathcal{P}}(\mathbf{S}_{\mathcal{P}}\cdot
\mathbf{L_{\mathcal{P}}})$ is $-5A_{\mathcal{P}}$ and these are listed in Table~\ref{Table I-a}.
\footnotesize
\begin{table}[tb]
\caption{\sf $S$ ($P$)- wave pentaquark states $\mathcal{P}_{X_{i}}$ ($\mathcal{P}_{Y_{i}}$)
and their spin- and orbital angular momentum quantum numbers. In the expressions for the masses of the
$\mathcal{P}_{Y_{i}}$ states,  $M_{\mathcal{P}_{X_{i}}}= M_0 + \Delta M_i$ with $i=1,...,5$.}
\label{Table I}
\begin{tabular}{|l|l|l||l|l|l|l|l|l|l|}
\hline
Label  &
 $|S_{\mathcal{Q}},S_{\mathcal{Q}^{\prime }}, S_{\bar{c}}, L_{\mathcal{P}}\; ; J^{P}\rangle _{i}
$ & Mass & Label & $|S_{\mathcal{Q}},S_{\mathcal{Q}^{\prime }}, S_{\bar{c}}, L_{\mathcal{P}%
}\;; J^{P}\rangle _{i}$ & Mass \\ \hline
$\mathcal{P}_{X_{1}}$ & $|0_{\mathcal{Q}},0_{\mathcal{Q}^{\prime }},\frac{1}{2}_{\bar{c}} , 0;%
\frac{1}{2}^{-}\rangle _{1}$ & $M_{0}+\Delta M_{1}$ & $\mathcal{P}_{Y_{1}}$
& $|0_{\mathcal{Q}},0_{\mathcal{Q}^{\prime }},\frac{1}{2}_{\bar{c}},1;\frac{1}{2}^{+}\rangle _{1}
$ & $M_{\mathcal{P}_{X_{1}}}-2A_{\mathcal{P}}+B_{\mathcal{P}}$ \\ \hline
$\mathcal{P}_{X_{2}}$ & $|0_{\mathcal{Q}},1_{\mathcal{Q}^{\prime }},\frac{1}{2}_{\bar{c}}, 0;%
\frac{1}{2}^{-}\rangle _{2}$ & $M_{0}+\Delta M_{2}$ & $\mathcal{P}_{Y_{2}}$
& $|0_{\mathcal{Q}},1_{\mathcal{Q}^{\prime }},\frac{1}{2}_{\bar{c}},1;\frac{1}{2}^{+}\rangle _{2}
$ & $M_{\mathcal{P}_{X_{2}}}-2A_{\mathcal{P}}+B_{\mathcal{P}}$ \\ \hline
$\mathcal{P}_{X_{3}}$ & $|1_{\mathcal{Q}},0_{\mathcal{Q}^{\prime }},\frac{1}{2}_{\bar{c}}, 0;%
\frac{1}{2}^{-}\rangle _{3}$ & $M_{0}+\Delta M_{3}$ & $\mathcal{P}_{Y_{3}}$
& $|1_{\mathcal{Q}},0_{\mathcal{Q}^{\prime }},\frac{1}{2}_{\bar{c}},1;\frac{1}{2}^{+}\rangle _{3}
$ & $M_{\mathcal{P}_{X_{3}}}-2A_{\mathcal{P}}+B_{\mathcal{P}}$ \\ \hline
$\mathcal{P}_{X_{4}}$ & $|1_{\mathcal{Q}},1_{\mathcal{Q}^{\prime }},\frac{1}{2}_{\bar{c}}, 0;%
\frac{1}{2}^{-}\rangle _{4}$ & $M_{0}+\Delta M_{4}$ & $\mathcal{P}_{Y_{4}}$
& $|1_{\mathcal{Q}},1_{\mathcal{Q}^{\prime }},\frac{1}{2}_{\bar{c}}, 1;\frac{1}{2}^{+}\rangle _{4}
$ & $M_{\mathcal{P}_{X_{4}}}-2A_{\mathcal{P}}+B_{\mathcal{P}}$ \\ \hline
$\mathcal{P}_{X_{5}}$ & $|1_{\mathcal{Q}},1_{\mathcal{Q}^{\prime }},\frac{1}{2}_{\bar{c}}, 0;%
\frac{1}{2}^{-}\rangle _{5}$ & $M_{0}+\Delta M_{5}$ & $\mathcal{P}_{Y_{5}}$ & $|1_{\mathcal{Q}},1_{\mathcal{Q}^{\prime }},\frac{1}{2}_{\bar{c}},1;\frac{1}{2}^{+}\rangle _{5}$ & $M_{\mathcal{P}%
_{X_{5}}}-2A_{\mathcal{P}}+B_{\mathcal{P}}$ \\ \hline\hline
\end{tabular}%
\end{table}
\normalsize
\begin{figure}[t]
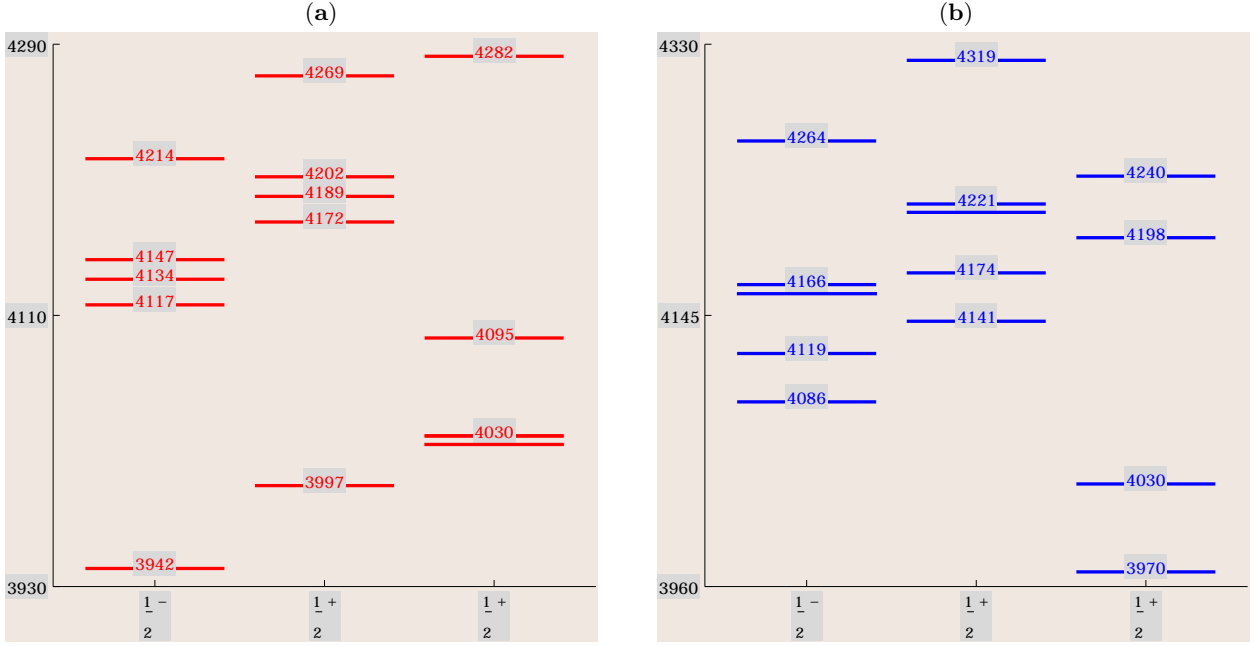

\begin{tabular}{cc}
\hspace{0.2cm}($\mathbf{a}$)&\hspace{0.4cm}($\mathbf{b}$) \\
      \includegraphics[scale=0.62]{figures/cqqqcbar-I} \ \ \ &  
\ \ \ \includegraphics[scale=0.62]{figures/cqqqcbar-II} 
\end{tabular}
\caption{\sf 
The mass spectrum (in~MeV) of the lowest $S$- and $P$-wave pentaquark states 
in the diquark-diquark-antiquark picture for the charmonium sector using 
the tetraquark type~I (a) and II (b) models having the quark flavor  $\bar{c}[cq][qq]$ ,
denoted as $c_1$  in Tab. \ref{Table III}. Note that the doubly drawn lines indicate 
states which are (almost) mass degenerate.
}
\label{Fig-Spectrum1}
\end{figure}

\begin{table}[tb]
\caption{\sf $S$ ($P$)- wave pentaquark states $\mathcal{P}_{X_{i}}$ ($\mathcal{P}_{Y_{i}}$)
and their spin- and orbital angular momentum quantum numbers. In the expressions for the masses of the
$\mathcal{P}_{Y_{i}}$ states, the terms $M_{\mathcal{P}_{X_{i}}}= M_0 + \Delta M_i$ with $i=6,\cdots , 9$. The $\Delta M_i$ are values of the $4 \times 4$ matrix given in~\cite{Ali:2016dkf} (c.f. Eq. (10)).}
\label{Table I-a}
\begin{tabular}{|l|l|l||l|l|l|l|l|l|}
\hline
Label  &
 $|S_{\mathcal{Q}},S_{\mathcal{Q}^{\prime }}, S_{\bar{c}}, L_{\mathcal{P}}\; ; J^{P}\rangle _{i}
$ & Mass & Label & $|S_{\mathcal{Q}},S_{\mathcal{Q}^{\prime }}, S_{\bar{c}}, L_{\mathcal{P}%
}\; ;J^{P}\rangle _{i}$ & Mass \\ \hline
$\mathcal{P}_{X_{6}}$ & $|0_{\mathcal{Q}},1_{\mathcal{Q}^{\prime }}, \frac{1}{2}_{\bar{c}} ,0;%
\frac{3}{2}^{-}\rangle _{6}$ & $M_{0}+\Delta M_{6}$ & $\mathcal{P}_{Y_{6}}$
& $|0_{\mathcal{Q}},1_{\mathcal{Q}^{\prime }}, \frac{1}{2}_{\bar{c}},1;\frac{1}{2}^{+}\rangle _{6}
$ & $M_{\mathcal{P}_{X_{6}}}-5A_{\mathcal{P}}+B_{\mathcal{P}}$ \\ \hline
$\mathcal{P}_{X_{7}}$ & $|1_{\mathcal{Q}},0_{\mathcal{Q}^{\prime }}, \frac{1}{2}_{\bar{c}}, 0;%
\frac{3}{2}^{-}\rangle _{7}$ & $M_{0}+\Delta M_{7}$ & $\mathcal{P}_{Y_{7}}$
& $|1_{\mathcal{Q}},0_{\mathcal{Q}^{\prime }},\frac{1}{2}_{\bar{c}}, 1;\frac{1}{2}^{+}\rangle _{7}
$ & $M_{\mathcal{P}_{X_{7}}}-5A_{\mathcal{P}}+B_{\mathcal{P}}$ \\ \hline
$\mathcal{P}_{X_{8}}$ & $|1_{\mathcal{Q}},1_{\mathcal{Q}^{\prime }},\frac{1}{2}_{\bar{c}}, 0;%
\frac{3}{2}^{-}\rangle _{8}$ & $M_{0}+\Delta M_{8}$ & $\mathcal{P}_{Y_{8}}$
& $|1_{\mathcal{Q}},1_{\mathcal{Q}^{\prime }},\frac{1}{2}_{\bar{c}}, 1;\frac{1}{2}^{+}\rangle _{8}
$ & $M_{\mathcal{P}_{X_{8}}}-5A_{\mathcal{P}}+B_{\mathcal{P}}$ \\ \hline
$\mathcal{P}_{X_{9}}$ & $|1_{\mathcal{Q}},1_{\mathcal{Q}^{\prime }},\frac{1}{2}_{\bar{c}}, 0;%
\frac{3}{2}^{-}\rangle _{9}$ & $M_{0}+\Delta M_{9}$ & $\mathcal{P}_{Y_{9}}$
& $|1_{\mathcal{Q}},1_{\mathcal{Q}^{\prime }},\frac{1}{2}_{\bar{c}}, 1;\frac{1}{2}^{+}\rangle _{9}
$ & $M_{\mathcal{P}_{X_{9}}}-5A_{\mathcal{P}}+B_{\mathcal{P}}$ \\ \hline
\end{tabular}%
\end{table}
\normalsize

%
\section{ $\boldsymbol{S}$- and $\boldsymbol{P}$-wave pentaquark spectrum with $\boldsymbol{J^{P} = \frac{1}{2}^{\pm}}$}\label{mass-spectrum}

In this section, we present the mass spectrum for the pentaquarks with a $\boldsymbol{c\bar{c}}$ 
and three light quarks, having $J^P=\frac{1}{2}^\pm$. The mass spectrum of some of the states with $J^{P} = \frac{1}{2}^{\pm}$ has already been calculated using QCD sum rules \cite{Wang-Huang, Wang-1509, penta-half}. The determination of the various input
parameters is explained in~\cite{Ali:2016dkf}  and for one particular case when the contents of pentaquark state is $\{\bar{c}\; [c q] \; [qq];  L_{\mathcal{P}}=0,1\}$ with $q = u\;, d$ the spectrum is shown in Fig.~\ref{Fig-Spectrum1}.
The five $S$-wave states have  $J^P=1/2^-$ (left-most group), and the two groups having  $J^P=1/2^+$ are the $P$-wave states, defined in Table~\ref{Table I} (shown in the middle of the frame) and in Table~\ref{Table I-a} (shown as the right-most group).

 For all the other possibilities of the light quark contents given in Table \ref{Table III}, the values of the estimated masses of the $J^{P}= {\frac{1}{2}^{\pm}}$ are presented in Tables~\ref{TableIV-1} - \ref{TableIV-3}. In order to calculate these spectra, the corresponding values of  $\Delta M_{i}$ for $i= 1, \cdots, 9$, obtained on diagonalizing the symmetric $5 \times 5$ matrix whose entries are given in Eqs.~(\ref{matrix-entires}) and (\ref{off-diag-matrix-entries}), for $i = 1, \cdots, 5$, and the $4 \times 4$ matrix given in~\cite{Ali:2016dkf} (c.f. Eq. (10)), for $i = 6, \cdots, 9$,  are mentioned in Table~\ref{TableM15}. 

\begin{table}[tb]
\caption{\sf Quark flavor contents of the pentaquarks  (with $q=u$ or $d$) arranged as $ \bar{c}$ and two
diquarks, and the corresponding flavor labels $c_i$ ($i=1,...,5$) used to characterize these states in the text.}
\label{Table III}
\begin{tabular}{|l|l|l|l|l|l|}
\hline
Quark contents &$\bar{c}[cq][qq]$ &$\bar{c}[cq][sq]$&
$\bar{c}[cs][qq]$ &$\bar{c}[cs][sq]$ &$\bar{c}[cq][ss]$ \\ 
\hline
Label &$\quad c_1$ &$\quad c_2$&$\quad c_3$ &$\quad c_4$ &$\quad c_5$\\
\hline\hline
\end{tabular}%
\end{table}

\begin{table*}[tbp]
\caption{\sf Numerical value of $\Delta M_{i}$ (in MeV) for $i = 1, \cdots , 9$ for the five different light quark
 combinations given in Table \ref{Table III}.}
\label{TableM15}
\begin{tabular}{|l|l|l|l|l|l|l|l|l|l|}
\hline
$\Delta M_{i} \,$& $\Delta M_{1}$ & $\Delta M_{2}$ & $\Delta M_{3}$ &$\Delta M_{4} $  &$\Delta M_{5}$ &
$\Delta M_{6}$ & $\Delta M_{7}$ & $\Delta M_{8}$ &$\Delta M_{9}$\\
\hline
$c_1$ & $-278 $ $(-126)$ & $-95$ $(-93)$ & $-78 $ $(-50)$ & $-65$ $(-46)$ & $2$ $(52)$  
& $-79$ $(-140)$ & $-79$ $(-79)$ & $-15 $ $(88)$ & $175$ $(130)$  \\ 
\hline
$c_2$ & $-224$ $(-98)$ & $-96$ $(-64)$ & $-55$ $(-60)$ & $-53$ $(-58)$ & $-24$ $(28)$ & 
$-77$ $(-161)$ & $-54$ $(-20)$ & $-1$ $(70)$ & $132$ $(111)$ \\ 
\hline
$c_3$ & $ -182$ $(-126)$ & $-96$ $(-93)$ & $-87$ $(-50)$ & $-40$ $(-46)$ & $-15$ $(52)$ &
$ -79$ $(-140)$ & $-54$ $(-79)$ & $-17$ $(88)$ & $134$ $(130)$ \\ 
\hline
$c_4$ & $-252$ $(-98)$ & $ -111$ $(-64)$ & $-49$ $(-60)$  &$-49$ $(-58)$ & $-16$ $(28)$ &
$-111$ $(-161)$ & $ -42$ $(-20)$ & $10$ $(70)$  &$144$ $(111)$\\ 
\hline
$c_5$ & $-186$ $(-144)$ & $-112$ $(-110)$ & $-97$ $(-45)$ & $-34$ $(-42)$ & $-4$ $(67)$
& $-112$ $(-131)$ & $-44$ $(-112)$ & $13$ $(102)$ & $145$ $(143)$\\ 
\hline\hline%
\end{tabular}%
\end{table*}
%
\begin{table*}[tbp]
\caption{\sf Masses of the hidden charm $S$-wave pentaquark states $\mathcal{P}_{X_i}$ (in MeV) formed through different diquark-diquark-anti-charm quark combinations in type I and type II models of tetraquarks. The masses given in the parentheses are for the input values taken from the type II model and the quoted errors are obtained from the uncertainties in the input parameters in the effective
Hamiltonian. }
\label{TableIV-1}
\begin{tabular}{|l|l|l|l|l|l|}
\hline
$\mathcal{P}_{X_{i}} \quad$& $\mathcal{P}_{X_{1}}$ & $\mathcal{P}_{X_{2}}$
& $\mathcal{P}_{X_{3}}$ &$\mathcal{P}_{X_{4}} $  &$\mathcal{P}_{X_{5}}
$ \\ 
\hline
$c_1$ & $3942\pm72$ $(4086 \pm 42)$ & $4117\pm42$ $(4119 \pm 42)$ &
$4134\pm38$ $(4162 \pm 38)$ & $4147\pm38$ $(4166\pm 38 )$ & $4214\pm45$ $(4264 \pm 41)$   \\ 
\hline
$c_2$ & $3967\pm55$ $(4094 \pm 44)$ & $4096\pm 46$ $(4128 \pm 44)$ & $ 4137\pm 44$ $(4132 \pm 43)$ & $ 4139\pm 43$ $(4134 \pm 42)$ & $4168\pm 44$ $(4220 \pm 43)$  \\ 
\hline
$c_3$ & $4262\pm 48$ $(4318 \pm 42)$  & $ 4348\pm 41$ $(4351 \pm 42)$ &$ 4357\pm 39$ $(4392\pm 38)$  & $4404\pm 39$ $(4398\pm 38)$ & $4429\pm 40$ $(4496 \pm 41)$ \\ 
\hline
$c_4$ & $4172\pm 48$ $(4326 \pm 44)$ & $ 4313\pm 44$ $(4360 \pm 43)$ & $ 4375\pm 43$ $(4364 \pm 43)$ & $4375\pm 43$ $(4366 \pm 43)$ & $4408\pm 44$ $(4452 \pm 43)$\\ 
\hline
$c_5$ & $4522\pm 51$ $(4564  \pm 44)$ & $ 4596\pm 44$ $(4598 \pm 44)$ & $4611\pm 43$ $(4662 \pm 43)$ & $4674\pm 43$ $(4666 \pm 43)$ & $4704\pm 43$ $(4775 \pm 44)$\\ \hline \hline%
\end{tabular}
\end{table*}%
\begin{table*}[tbp]
\caption{\sf Masses of the hidden charm $P$-wave pentaquark states $\mathcal{P}_{Y_i}$ (in MeV) formed through different diquark-diquark-anti-charm quark combinations in type I and type II models of tetraquarks. The masses given in the parentheses are for the input values taken from the type II model and the quoted errors are obtained from the uncertainties in the input parameters in the effective
Hamiltonian.}
\label{TableIV-2}
\begin{tabular}{|l|l|l|l|l|l|}
\hline
$\mathcal{P}_{Y_{i}} \quad$& $\mathcal{P}_{Y_{1}}$ & $\mathcal{P}_{Y_{2}}$
& $\mathcal{P}_{Y_{3}}$ &$\mathcal{P}_{Y_{4}} $  &$\mathcal{P}_{Y_{5}}$ \\ 
\hline
$c_1$ & $3997\pm 73 $ $(4141 \pm 44)$ & $4172\pm 44$ $(4174 \pm 44)$ & $4189\pm 40 $ $(4217 \pm 40)$ & $4202\pm 39$ $(4221 \pm 40)$ & $4269\pm 47$ $(4319 \pm 43)$  \\ 
\hline
$c_2$ & $4023 \pm 56$ $(4149 \pm 45)$ & $4151\pm 47$ $(4183 \pm 45)$ & $4192\pm 45$ $(4187 \pm 44)$ & $4194\pm 45$ $(4189 \pm 44)$ & $4223\pm 46$ $(4275 \pm 45)$  \\ 
\hline
$c_3$ & $ 4317\pm 50$ $(4373 \pm 44)$ & $4403\pm 43$ $(4406 \pm 44)$ & $4412\pm 41$ $(4449  \pm 40)$ & $4459 \pm 41$ $(4453 \pm 40)$ & $4484 \pm 42$ $(4551 \pm 43)$  \\ 
\hline
$c_4$ & $4227\pm 50$ $(4381 \pm 45)$ & $ 4368\pm 45$ $(4415  \pm 45)$ & $ 4430\pm 45$ $(4419 \pm 44)$  &$4430\pm 45$ $(4421 \pm 44)$ & $4463\pm 45$ $(4507 \pm 45)$ \\ 
\hline
$c_5$ & $4577\pm 52$ $(4619 \pm 45)$ & $ 4651\pm 45$ $(4653 \pm 45)$ & $4666\pm 44$ $(4717 \pm 44)$ & $4729\pm 44$ $(4721 \pm 44)$ & $4759\pm 45$ $(4830 \pm 45)$\\ 
\hline\hline%
\end{tabular}%
\end{table*}
%
\begin{table*}[tbp]
\caption{\sf Masses of the hidden charm $P$-wave pentaquark states $\mathcal{P}_{Y_i}$ (in MeV)  formed through different diquark-diquark-anti-charm quark combinations in type I and type II models of tetraquarks. These states are obtained by combining the spin $\frac{3}{2}$ of pentaquark with $L=1$ to have the final pentaquark states with $J^{P}= \frac{1}{2}^{+}$. The masses given in the parentheses are for the input values taken from the type II model and the quoted errors are obtained from the uncertainties in the input parameters in the effective
Hamiltonian.}
\label{TableIV-3}
\begin{tabular}{|l|l|l|l|l|l|}
\hline
$\mathcal{P}_{Y_{i}} \quad$& $\mathcal{P}_{Y_{6}}$ & $\mathcal{P}_{Y_{7}}$
& $\mathcal{P}_{Y_{8}}$ &$\mathcal{P}_{Y_{9}} $ \\ 
\hline
$c_1$ & $4030\pm 62 $ $(3970 \pm 50)$ & $4030\pm 62$ $(4030 \pm 62)$ & $4095\pm 63 $ $(4198 \pm 50)$ & $4282\pm 63$ $(4240 \pm 50)$  \\ 
\hline
$c_2$ & $4012 \pm65$ $(3929 \pm 53)$ & $4036\pm 56$ $(4069 \pm 56)$ & $4088\pm 61$ $(4159 \pm 53)$ & $4222\pm 56$ $(4201 \pm 52)$   \\ 
\hline
$c_3$ & $ 4263\pm 62$ $(4202 \pm 50)$ & $4288\pm 52$ $(4262 \pm 63)$ & $4341\pm 57$ $(4430  \pm 50)$ & $4475 \pm 52$ $(4472 \pm 50)$  \\ 
\hline
$c_4$ & $4210\pm 56$ $(4161 \pm 53)$ & $ 4279\pm 54$ $(4301  \pm 56)$ & $ 4332\pm 60$ $(4391 \pm 53)$  &$4465\pm 55$ $(4433 \pm 52)$ \\ 
\hline
$c_5$ & $4493\pm 56$ $(4474 \pm 53)$ & $ 4561\pm 54$ $(4493 \pm 56)$ & $4618\pm 59$ $(4707 \pm 53)$ & $4750\pm 55$ $(4748 \pm 52)$\\ 
\hline\hline%
\end{tabular}%
\end{table*}

\section{Production of $\boldsymbol{J^{P}(\frac{1}{2}^{\pm})}$ pentaquark states in the weak decays of the $\boldsymbol{b}$-baryons}\label{production-baryon}
The possible production of these charmed pentaquark states is possible through the weak decays of $b$-baryon. The effective weak Hamiltonian inducing $b \to c \bar{c} q$ transition:
\begin{equation}
H^W_{\text{eff}} = \frac{4 G_F}{\sqrt 2} 
\sum_{q = d, s} V_{cb} V_{cq}^* \left ( 
C_1 \, O_1^{(q)} + C_2  \, O_2^{(q)} \right ) ,
\label{weak-hamiltonian}
\end{equation}
with $q$ being $s$ and $d$ quarks correspond to the Cabibbo-allowed $\Delta I = 0$, $\Delta S = -1$ and the Cabibbo-suppressed 
$\Delta I = -1/2$, $\Delta S = 0$ transitions, respectively. In Eq. (\ref{weak-hamiltonian}), $G_F$ is the Fermi coupling constant,~$V_{ij}$ are the CKM matrix 
elements, and~$C_i$ are the Wilson coefficients of the four-quark 
operators~$O_i^{(q)}$ ($q=d,\; s$), defined as
\begin{equation}
O_1^{(q)} = (\bar q_\alpha c_\beta)_{V-A} (\bar c_\alpha b_\beta)_{V-A}, \;\;\;
O_2^{(q)} = (\bar q_\alpha c_\alpha)_{V-A} (\bar c_\beta b_\beta)_{V-A},
\label{tree-operators}
\end{equation}
where~ 
$(\bar{q}_\alpha q^\prime_\beta)_{V-A} = 
 [\bar{q}_\alpha \gamma_\mu \left ( 1 - \gamma_5 \right ) q^\prime_\beta]$ 
are the left handed charged currents, $\alpha$ and ~$\beta$ are $SU(3)_C$ color indices.

The amplitude corresponding to the decay of $b$-baryon from the flavor anti-triplet and sextet according to the $SU(3)_F$-group, denoted by $\mathcal{B}$ and ~$\mathcal{C}$, respectively, into pentaquark state from the $SU(3)_F$ octet $(\mathcal{P})$ along with a light pseudoscalar meson $(\mathcal{M})$ can be written as
\begin{equation}
\mathcal{A} = \left\langle \mathcal{P M} 
\left \vert H^W_{\text{eff}} \right \vert 
\mathcal{B} (\mathcal{C}) \right\rangle,
\label{production-amplitude}
\end{equation}
where $ H^W_{\text{eff}}$ is defined in Eq. (\ref{weak-hamiltonian}). In (\ref{production-amplitude}), $\mathcal{B}$ is a flavor
anti-triplet $b$-baryon with the light-quark spin $j^P=0^+$. The explicit expressions of $\mathcal{B}$, 
a light pseudoscalar meson in the $SU(3)_F$ octet $\mathcal{M}$ and the final state pentaquark 
 $\mathcal{P}$ are given in \cite{Ali:2016dkf}.\\
 
 \vspace*{2mm}
In the limit of heavy quark symmetry, the tree amplitudes for the anti-triplet 
$b$-baryon decays into an octet pentaquark and a pseudoscalar meson 
can be decomposed as follows ($q = d$ or~$s$) \cite{Ali:2016dkf}:
\begin{eqnarray}
\mathcal{A}^J_{t8} \left ( q \right ) & = & 
T^J_3 \left\langle \mathcal{P}_k^i \mathcal{M}_j^k 
\left\vert H \left ( \bar 3 \right )^j \right\vert 
\mathcal{B}_{l m} \right\rangle 
\varepsilon^{i l m} + 
T^J_5 \left\langle \mathcal{P}_{j^\prime}^{l} \mathcal{M}_j^i 
\left\vert H \left ( \bar 3 \right )^j \right\vert 
\mathcal{B}_{m j^\prime} \right\rangle 
\varepsilon_{i l m}, \label{triplet-amplitude}
\end{eqnarray} 
where the superscript~$J$ represents the spin of the final-state 
pentaquark, $J = \frac{1}{2}$.  The Feynman diagrams corresponding to the above amplitudes are shown in Fig. 7 \cite{Ali:2016dkf}. 
Similarly, in the case of  the sextet $b$-baryons 
from $\mathcal{C}_{ij} (6)$ decaying into decuplet pentaquark states 
from~$\mathcal{P}_{i j k}$, the decay amplitude in the heavy quark symmetry approximation can be written in the form ($q = d$ or~$s$):
\begin{equation}
\mathcal{A}^J_{t10} \left ( q \right ) = T_5^{s\, J} 
\left\langle \mathcal{P}_{k j^\prime m} \mathcal{M}_l^k 
\left\vert H \left ( \bar 3 \right )^l \right\vert 
\mathcal{C}_{m j^\prime} \right\rangle .
\label{sextet-amplitude} 
\end{equation}%
where $\mathcal{C}_{ij}\left( 6\right)$
and decuplet $\mathcal{P}_{i j k}$ (symmetric in all indices) are listed in \cite{Ali:2016dkf}. 

With these amplitudes, the estimates of the ratios of decay widths $\Gamma(\mathcal{B}(\mathcal{C}) \to \mathcal{P}^{1/2^{\pm}}\mathcal{M})/\Gamma(\Lambda^{0}_b \to P_{p}^{{\{Y_2\}}_{c_1}}K^{-})$, where $P_{p}^{{\{Y_2\}}_{c_1}}$ is the measured state with the mass $4450$ MeV and $J^{P} = \frac{5}{2}^{+}$, are given in Tables \ref{Relative-Rates11} - \ref{Relative-Rates14}.
 \begin{table*}[tbp]
\caption{\sf {Estimate of the ratios of decay widths $\Gamma(\mathcal{B}(\mathcal{C}) \to \mathcal{P}^{1/2^{-}}\mathcal{M})/\Gamma(\Lambda^{0}_b \to P_{p}^{{\{Y_2\}}_{c_1}}K^{-})$ for $\Delta S = 1$ transitions by using the masses of pentaquark states  $\mathcal{P}^{c_i}_{X_3}$ worked out in this work.  Here $P_{p}^{{\{Y_2\}}_{c_1}}$ is the state with mass $4450$ MeV and $J^{P} = \frac{5}{2}^{+}$ that has been measured recently at the LHCb  \cite{Aaij:2015tga}.}}
\label{Relative-Rates11}
\begin{tabular}{|c|c|c|c|c|}
\hline
Decay Process & $\Gamma/\Gamma(\Lambda^{0}_b \to P_{p}^{{\{Y_2\}}_{c_1}}K^{-})$ & Decay Process & $\Gamma/\Gamma(\Lambda^{0}_b \to P_{p}^{{\{Y_2\}}_{c_1}}K^{-})$ \\ 
\hline
$\Lambda _{b}\rightarrow P_{p}^{\{X_{3}\}_{c_{1}}}K^{-}$ & $%
0.61$  &  $\Xi _{b}^{-}\rightarrow P_{\Sigma ^{-}}^{\{X_{3}\}_{c_{2}}}  \bar{K}^{0}$
& $0.83 $   \\ 
\hline
$\Lambda _{b}\rightarrow P_{n}^{\{X_{3}\}_{c_{1}}}\bar{K}^{0}$ & $%
0.61 $ & $\Xi _{b}^{0}\rightarrow P_{\Sigma ^{+}}^{\{X_{3}\}_{c_{2}}} K^{-}$ & $%
0.83$ \\ 
\hline
$\Lambda _{b}\rightarrow P_{\Lambda^{0}}^{\{X_{3}\}_{c_{3}}} \eta^{\prime}$ 
& $0.04 $  & $\Lambda _{b}\rightarrow P_{\Lambda^{0}}^{\{X_{3}\}_{c_{3}}} \eta$ 
& $0.08$ \\ 
\hline
 $\Xi _{b}^{-}\rightarrow P_{\Sigma^{0}}^{\{X_{3}\}_{c_{2}}}  K^{-}$ & $%
1$ & $\Xi _{b}^{-}\rightarrow P_{\Lambda ^{0}}^{\{X_{3}\}_{c_{2}}}  K^{-}$ & $%
0.10$  \\ 
\hline
$\Omega _{b}^{-}\rightarrow P_{\Xi_{10} ^{-}}^{\{X_{4}\}_{c_{5}}} \bar{K}^{0}$
& $0.3$  & $\Omega _{b}^{-}\rightarrow P_{\Xi_{10} ^{0}}^{\{X_{4}\}_{c_{5}}} K^{-}$ & $%
0.3$  \\
\hline \hline%
\end{tabular}%
\end{table*}

\begin{table*}[tbp]
\caption{\sf {Estimate of the ratios of the decay widths  $\Gamma(\mathcal{B}(\mathcal{C}) \to \mathcal{P}^{1/2^{-}}\mathcal{M})/\Gamma(\Lambda^{0}_b \to P_{p}^{{\{Y_2\}}_{c_1}}K^{-})$, where $P_{p}^{{\{Y_2\}}_{c_1}}$ is a state with mass 4450 MeV that has recently been measured at LHCb  \cite{Aaij:2015tga}, for $\Delta S = 0$ transitions. In comparison to the  $\Delta S = 1$ transitions, these transitions are suppressed by a factor $|V^{\ast}_{cd}/V^{\ast}_{cs}|^2$. 
Other input values are the same as in Table \ref{Relative-Rates11}. }}
\label{Relative-Rates12}
\begin{tabular}{|c|c|c|c|c|}
\hline
Decay Process & $\Gamma/\Gamma(\Lambda^{0}_b \to P_{p}^{{\{Y_2\}}_{c_1}}K^{-})$ & Decay Process & $\Gamma/\Gamma(\Lambda^{0}_b \to P_{p}^{{\{Y_2\}}_{c_1}}K^{-})$ \\ 
\hline
$\Lambda _{b}\rightarrow P_{p}^{\{X_{3}\}_{c_{1}}} \pi ^{-}$  & $%
0.04$ &$\Lambda _{b}\rightarrow P_{n}^{\{X_{3}\}_{c_{1}}} \pi ^{0}$ &$0.02  $   \\ 
\hline
$\Lambda _{b}\rightarrow P_{n}^{\{X_{3}\}_{c_{1}}} \eta$  & $%
0.01$ &$\Lambda _{b}\rightarrow P_{n}^{\{X_{3}\}_{c_{1}}} \eta^{\prime}$ &$0 $   \\ 
\hline
 $\Xi _{b}^{-}\rightarrow P_{\Xi ^{-}}^{\{X_{3}\}_{c_{4}}} K^{0}$ & $0.03$ & $\Xi _{b}^{-}\rightarrow P_{\Sigma ^{0}}^{\{X_{3}\}_{c_{2}}}  \pi ^{-}$ & $0.01$ \\ 
\hline
 $\Xi _{b}^{-}\rightarrow P_{\Sigma ^{-}}^{\{X_{3}\}_{c_{2}}} \eta$ &$0.01$ &$\Xi _{b}^{-}\rightarrow P_{\Sigma ^{-}}^{\{X_{3}\}_{c_{2}}} \eta^{\prime}$ &$0.01 $ \\ 
\hline
$\Xi _{b}^{-}\rightarrow P_{\Sigma ^{-}}^{\{X_{3}\}_{c_{2}}} \pi ^{0}$ & $%
0.02$  & $\Xi _{b}^{0}\rightarrow P_{\Sigma ^{0}}^{\{X_{3}\}_{c_{2}}} \pi ^{0}$ & $%
0.01$  \\ 
\hline
$\Xi _{b}^{0}\rightarrow P_{\Lambda ^{0}}^{\{X_{3}\}_{c_{2}}} \eta$  & $0$ & $\Xi _{b}^{0}\rightarrow P_{\Lambda ^{0}}^{\{X_{3}\}_{c_{2}}} \eta^{\prime}$ & $%
0$  \\
\hline
$\Xi _{b}^{0}\rightarrow P_{\Lambda ^{0}}^{\{X_{3}\}_{c_{2}}} \pi ^{0}$  & $0.01$ & $\Omega _{b}^{-}\rightarrow P_{\Xi_{10} ^{-}}^{\{X_{3}\}_{c_{5}}} \pi ^{0}$ & $%
0.01$  \\
\hline
$\Omega _{b}^{-}\rightarrow P_{\Xi_{10}^{0}}^{\{X_{4}\}_{c_{5}}} \pi ^{-}$ & $%
0.02 $ & &\\ \hline \hline%
\end{tabular}%
\end{table*}

 \begin{table*}[tbp]
\caption{\sf {Estimate of the ratios of decay widths $\Gamma(\mathcal{B}(\mathcal{C}) \to \mathcal{P}^{1/2^{+}}\mathcal{M})/\Gamma(\Lambda^{0}_b \to P_{p}^{{\{Y_2\}}_{c_1}}K^{-})$ for $\Delta S = 1$ transitions by using the masses of pentaquark states  $\mathcal{P}^{c_i}_{Y_3}$ worked out in this work.  Here $P_{p}^{{\{Y_2\}}_{c_1}}$ is the state with mass $4450$ MeV and $J^{P} = \frac{5}{2}^{+}$ that has been measured recently at the LHCb  \cite{Aaij:2015tga}.}}
\label{Relative-Rates13}
\begin{tabular}{|c|c|c|c|c|}
\hline
Decay Process & $\Gamma/\Gamma(\Lambda^{0}_b \to P_{p}^{{\{Y_2\}}_{c_1}}K^{-})$ & Decay Process & $\Gamma/\Gamma(\Lambda^{0}_b \to P_{p}^{{\{Y_2\}}_{c_1}}K^{-})$ \\ 
\hline
$\Lambda _{b}\rightarrow P_{p}^{\{X_{3}\}_{c_{1}}}K^{-}$ & $%
0.39$  &  $\Xi _{b}^{-}\rightarrow P_{\Sigma ^{-}}^{\{X_{3}\}_{c_{2}}}  \bar{K}^{0}$
& $0.43 $   \\ 
\hline
$\Lambda _{b}\rightarrow P_{n}^{\{X_{3}\}_{c_{1}}}\bar{K}^{0}$ & $%
0.39 $ & $\Xi _{b}^{0}\rightarrow P_{\Sigma ^{+}}^{\{X_{3}\}_{c_{2}}} K^{-}$ & $%
0.43$ \\ 
\hline
$\Lambda _{b}\rightarrow P_{\Lambda^{0}}^{\{X_{3}\}_{c_{3}}} \eta^{\prime}$ 
& $0.07 $  & $\Lambda _{b}\rightarrow P_{\Lambda^{0}}^{\{X_{3}\}_{c_{3}}} \eta$ 
& $0.07$ \\ 
\hline
 $\Xi _{b}^{-}\rightarrow P_{\Sigma^{0}}^{\{X_{3}\}_{c_{2}}}  K^{-}$ & $%
0.3$ & $\Xi _{b}^{-}\rightarrow P_{\Lambda ^{0}}^{\{X_{3}\}_{c_{2}}}  K^{-}$ & $%
0.1$  \\ 
\hline
$\Omega _{b}^{-}\rightarrow P_{\Xi_{10} ^{-}}^{\{X_{4}\}_{c_{5}}} \bar{K}^{0}$
& $0.29$  & $\Omega _{b}^{-}\rightarrow P_{\Xi_{10} ^{0}}^{\{X_{4}\}_{c_{5}}} K^{-}$ & $%
0.29$  \\
\hline \hline%
\end{tabular}%
\end{table*}

\begin{table*}[tbp]
\caption{\sf {Estimate of the ratios of the decay widths  $\Gamma(\mathcal{B}(\mathcal{C}) \to \mathcal{P}^{1/2^{+}}\mathcal{M})/\Gamma(\Lambda^{0}_b \to P_{p}^{{\{Y_2\}}_{c_1}}K^{-})$, where $P_{p}^{{\{Y_2\}}_{c_1}}$ is a state with mass 4450 MeV that has recently been measured at LHCb \cite{Aaij:2015tga}, for $\Delta S = 0$ transitions. In comparison to the $\Delta S = 1$ transitions, these transitions are suppressed by a factor $|V^{\ast}_{cd}/V^{\ast}_{cs}|^2$. 
Other input values are the same as in Table \ref{Relative-Rates13}. }}
\label{Relative-Rates14}
\begin{tabular}{|c|c|c|c|c|}
\hline
Decay Process & $\Gamma/\Gamma(\Lambda^{0}_b \to P_{p}^{{\{Y_2\}}_{c_1}}K^{-})$ & Decay Process & $\Gamma/\Gamma(\Lambda^{0}_b \to P_{p}^{{\{Y_2\}}_{c_1}}K^{-})$ \\ 
\hline
$\Lambda _{b}\rightarrow P_{p}^{\{X_{3}\}_{c_{1}}} \pi ^{-}$  & $%
0.02$ &$\Lambda _{b}\rightarrow P_{n}^{\{X_{3}\}_{c_{1}}} \pi ^{0}$ &$0.02  $   \\ 
\hline
$\Lambda _{b}\rightarrow P_{n}^{\{X_{3}\}_{c_{1}}} \eta$  & $%
0.01$ &$\Lambda _{b}\rightarrow P_{n}^{\{X_{3}\}_{c_{1}}} \eta^{\prime}$ &$0 $   \\ 
\hline
 $\Xi _{b}^{-}\rightarrow P_{\Xi ^{-}}^{\{X_{3}\}_{c_{4}}} K^{0}$ & $0.02$ & $\Xi _{b}^{-}\rightarrow P_{\Sigma ^{0}}^{\{X_{3}\}_{c_{2}}}  \pi ^{-}$ & $0.01$ \\ 
\hline
 $\Xi _{b}^{-}\rightarrow P_{\Sigma ^{-}}^{\{X_{3}\}_{c_{2}}} \eta$ &$0.01$ &$\Xi _{b}^{-}\rightarrow P_{\Sigma ^{-}}^{\{X_{3}\}_{c_{2}}} \eta^{\prime}$ &$0 $ \\ 
\hline
$\Xi _{b}^{-}\rightarrow P_{\Sigma ^{-}}^{\{X_{3}\}_{c_{2}}} \pi ^{0}$ & $%
0.01$  & $\Xi _{b}^{0}\rightarrow P_{\Sigma ^{0}}^{\{X_{3}\}_{c_{2}}} \pi ^{0}$ & $%
0.01$  \\ 
\hline
$\Xi _{b}^{0}\rightarrow P_{\Lambda ^{0}}^{\{X_{3}\}_{c_{2}}} \eta$  & $0$ & $\Xi _{b}^{0}\rightarrow P_{\Lambda ^{0}}^{\{X_{3}\}_{c_{2}}} \eta^{\prime}$ & $%
0$  \\
\hline
$\Xi _{b}^{0}\rightarrow P_{\Lambda ^{0}}^{\{X_{3}\}_{c_{2}}} \pi ^{0}$  & $0$ & $\Omega _{b}^{-}\rightarrow P_{\Xi_{10} ^{-}}^{\{X_{3}\}_{c_{5}}} \pi ^{0}$ & $%
0.00$  \\
\hline
$\Omega _{b}^{-}\rightarrow P_{\Xi_{10}^{0}}^{\{X_{4}\}_{c_{5}}} \pi ^{-}$ & $%
0.01 $ & &\\ \hline \hline%
\end{tabular}%
\end{table*}

\section{Decays of $\boldmath{J = \frac{1}{2}}$ pentaquark states and corresponding thresholds}

The  pentaquark states discussed here can be produced through the decays of the $b$-baryons, and they decay further into stable baryons and mesons. The mass of the $J = \frac{1}{2}$ pentaquark state having the flavor content of a proton $(p)$ and a $J/\psi$, denoted by $P_{p}^{\{X_{3}\}_{c_{1}}}$, is estimated as $4134\pm 38$ MeV. The error  arise from the ranges of the different input parameters in the effective Hamiltonian framework, assuming that the parameters subsume the essential underlying physics. Within this framework, the nominal mass of $P_{p}(4134)$ is about $100$ MeV above the $J/\psi\,p$ threshold,
 $4035$ MeV \cite{Olive:2016xmw}.  If the mass of $P_{p}^{\{X_{3}\}_{c_{1}}}$ is indeed higher than the $J/\psi\,p$ threshold, 
 as anticipated here, then it can be observed in the future by  the LHCb in the $J/\psi\;p$ channel.  
If not, this state may lie just below the  $J/\psi\;p$ threshold, and
 the channel to look for is the $(J/\psi)^{\ast}\,p$ where the virtual $(J/\psi)^{\ast}$ decays, among other
 states,  to the dileptons  $\mu^{+}\mu^{-}$ and $e^+e^-  $. There is another possible decay of the state $P_{p}^{\{X_{3}\}_{c_{1}}}$,  $P_{p}^{\{X_{3}\}_{c_{1}}}(4134) \to \eta_{c}\,p$,
 which has the threshold  of $3963$ MeV \cite{Olive:2016xmw}.

Similarly, in the case of the pentaquark $P_{\Lambda^{0}(\Sigma^0)}^{\{X_{3}\}_{c_{2}}}$, which could be produced in the decays of $\Xi_{b}^{-}$ along with a $\pi^{-}$, the mass estimated here is $4137\pm 44$ MeV.
The nominal mass of this state  is below the corresponding threshold for $J/\psi\,\Lambda^{0}(\Sigma^{0})$ which is around $4211\;(4290)$ MeV \cite{Olive:2016xmw}. 
Thus, the final states to look for are $(J/\psi)^*\,\Lambda^{0}(\Sigma^{0})$, where the
virtual $(J/\psi)^*$ is measured in the  $\mu^{+}\mu^{-}$ and $e^+e^-  $ modes.
  However, the state with the same overall quark flavor quantum numbers $P_{\Lambda^{0}(\Sigma^0)}^{\{X_{3}\}_{c_{3}}}$ (see Table \ref{Table III}) is estimated to have  the mass $4357 \pm 39$ MeV. This can be produced along with
an  $\eta_{8}$ in the decays of $\Lambda_{b}^{0}$ through $\Delta S = 1$ transitions.
  This lies above the $J/\psi\,\Lambda^{0}(\Sigma^{0})$  threshold by almost $120\;(70)$ MeV \cite{Olive:2016xmw}. Therefore, there is a possibility to observe this state in the $\Lambda_{b}^{0}$ decays. Again, if the mass of this state turns out to be
  below the stated thresholds, then the search decay modes are  
$P_c^{+}(J^P=1/2^{\pm}) \to \eta_c\; \Lambda^{0}(\Sigma^0)\;, \mu^+ \mu^- \; \Lambda^{0}(\Sigma^0),\;
e^+ e^- \; \Lambda^{0}(\Sigma^0)$.

In the case of the decays $\Omega _{b}^{-}\rightarrow P_{\Xi_{10} ^{-}}^{\{X_{3}\}_{c_{5}}} \pi^ {0}$ and $\Omega _{b}^{-}\rightarrow P_{\Xi_{10} ^{0}}^{\{X_{4}\}_{c_{5}}}K^ {-}$, which are $\Delta S = 0$ and $\Delta S =1$ transitions, respectively, the masses of the $P_{\Xi_{10} ^{0}}^{\{X_{4}\}_{c_{5}}}$ and $P_{\Xi_{10} ^{-}}^{\{X_{3}\}_{c_{5}}}$ are estimated to be significantly above the $(\Xi_{10}\;J/\psi)$-threshold. Thus, there exist exciting possibilities to search
for spin-$1/2$ pentaquarks in various $b$-baryon decays at the LHCb.

\textbf{Acknowledgments}

One of us (A.A.) would like to thank Luciano Maiani, Antonello Polosa and Sheldon Stone for helpful discussions.

%
\end{document}